\documentstyle[12pt,epsf]{article}
\begin{document}
\vspace*{0.5 true in}
\begin{center}
{\bf Level density and level density parameter in medium heavy}\\
{\bf nuclei including thermal and quantal fluctuation effects}\\
\vspace*{0.5 true in}
B. K. Agrawal and A. Ansari\\

Institute of Physics, Bhubaneswar 751 005, India
\end{center}

\vspace*{0.7 true in}
\begin{abstract}
   In a {\it realistic} application of the SPA + RPA theory for calculation of
the nuclear level densities we find that quantal fluctuation corrections
(RPA) are important even up to temperature $T = 2.0$ $MeV$.
This leads to  
 a good agreement 
between  calculated numbers and the available
experimental data for $^{104}Pd$ and $^{114}Sn$, particularly the
excitation energy ($E^*$) dependence. Furthermore, we also argue that
 $a=S^2/4E^*$ is the only correct definition of the level density parameter
in the present context which is also consistent with the Bethe like
level density formula. 
\end{abstract}
\newpage
In the last a few years there have been considerable efforts to develop microscopic 
methods for the calculation of accurate values of level densities as a function
of excitation
energies\cite{Chang anp66}-\cite{puddu anp206}. One of these methods which is of present interest is based
on the auxiliary field  path integral representation of the partition
function for a given nuclear Hamiltonian. The path integral representation
can be obtained in two ways: (a) the so called as shell model Monte Carlo
(SMMC) method\cite{{Lang prc48},{kdl pr278}}
and (b) the SPA$+$RPA approach\cite{puddu anp206}
 which includes the thermal fluctuations 
through static path approximation (SPA) and the quantal fluctuations
about static paths are included using random phase approximaton (RPA). 
It is now well established that SMMC approach can be applied even at
very low temperatures. On the otherhand, SPA$+$RPA approach is
computationally faster than SMMC approach and can be used for moderately low
to high temperatures, typically, for $T \, \geq\, 0.2$ $MeV$.
However, in most of these studies so far the main emphasis has been
to demonstrate, through simple model studies, 
  the relative defferences between the SPA and
SPA$+$RPA results for the temperature dependence of the energy and the 
level density.

Recently\cite{drc prc}, the level densities  as well as the level density
parameters have been extracted for the medium mass nuclei 
$^{104}Pd$ through the measurement of proton yields 
in reaction $^{93}Nb(^{12}C,p)^{104}Pd$ and that of $^{114}Sn$
through $^{103}Rh(^{12}C,p)^{114}Sn$ reaction.
 These data are available for 
the excitation energy ($E^*$) ranging from $5 - 25$ $MeV$ or equivalently
$T\simeq 0.5 - 1.5$ $MeV$ which is well suited to test the feasibility
of the SPA$+$RPA approach. In this letter we calculate the level density
and the level density parameter as a function of $E^*$ for $^{104}Pd$
and $^{114}Sn$ using SPA$+$RPA approach with a quadrupole-quadrupole 
interaction model Hamiltonian
\begin{equation}
H=H_0-\frac{1}{2}\chi\sum_{\mu=-2}^{2}(Q_\mu)^2.
\label{H}
\end{equation}
In the above, $H_0$ represents the  spherical part, $Q_0=
Q'_0$, $Q_{+\mu}=\frac{1}{\sqrt {2}} (Q'_\mu+Q'^\dagger_\mu)$ and
$Q_{-\mu}=\frac{i}{\sqrt{2}}(Q'_\mu-Q'^\dagger_\mu)$ with $\mu =$ 1 and
2 and $Q's$ stand for the usual quadrupole moment operators. 
Value of the quadrupole interaction strength $\chi=120 A^{-5/3} f_c$ $MeV$
($A$ denotes the mass number) is taken from Ref. \cite{aberg plb} where 
$f_c = 1 - 2 $ is a core polarization factor. 

The grand canonical partition function in SPA$+$RPA takes the following 
form \cite{Lauritzen plb246}
\begin{eqnarray}
{\cal Z}_{RPA}=4\pi^2\left (\frac{\alpha}{2\pi T}\right )^{5/2}
\int \beta^4 d\beta \int \mid sin 3\gamma\mid d\gamma 
 e^{-\frac{\alpha\beta^2}{2T}}\nonumber\\
\times  Tr\left [ e^{-H'/T} \right ] {\cal C}_{\scriptstyle {RPA}}.
\label{zrpa}
\end{eqnarray}
where, $\alpha=(\hbar\omega_0)^2/\chi$ with $\hbar\omega_0=41 A^{-1/3}$ $MeV$.
The quantities $H'$ and ${\cal C}_{
\scriptstyle {RPA}}$ are the single-particle Hamiltonian and the RPA correction
factor, respectively, given by
\begin{equation}
H'=H_0-\hbar\omega_0 \beta \left (Q_0\,cos\gamma + Q_{+2}\, sin\gamma 
\right )
\label{H'}
\end{equation}
and
\begin{equation}
{\cal C}_{\scriptstyle {RPA}}=\left (\prod_{m\not = 0}^{N_m}
 Det \mid C^m\mid \right ]^{-1}
\label{crpa}
\end{equation}
with
\begin{equation}
\label{cm}
C^m_{\mu\nu}=\delta_{\mu\nu} + \chi\sum_{ij}
\frac{\langle i\mid 
Q_\mu\mid j\rangle \langle j\mid Q_\nu \mid i\rangle}
 {\Delta_{ij}^2 + (2\pi m T)^2}
f_{ij}\Delta_{ij}
\end{equation}
where, $f_{ij}=f_i-f_j$ and $\Delta_{ij}=\epsilon_i-\epsilon_j$ with
$f_i$ being the Fermi distribution function and $\epsilon_i$ is the 
eigenvalue of $H'$. In the above,  $\mid i \rangle$ represents 
an eigenstate of $H'$ . The grand canonical trace
in eq. (\ref{zrpa}) can  simply be performed using
\begin{eqnarray}
Tr e^{-\beta' H'}=\left (
\prod_i[1+e^{-\beta'\epsilon_i+\alpha_p}]\right )
\left (\prod_j[1+e^{-\beta'\epsilon_j+\alpha_n}]\right )
\end{eqnarray}
where, $\beta'=1/T$ and $\alpha_p(\alpha_n)$ is the Lagrange multiplier
required to adjust the proton(neutron) numbers. 

The SPA representation of the partition function can be obtained by
putting ${\cal C}_{\scriptstyle {RPA}} = 1$. It is, therefore, 
clear from eqs.  (\ref{crpa},\ref{cm}) that for higher temperatures ${\cal
C}_{\scriptstyle {RPA}}
\longrightarrow 1$ or in other words $Z_{\scriptstyle {RPA}}
\longrightarrow Z_{\scriptstyle{SPA}}$.

Once the partition function is known, the level densiy or more precisely
the state density $W(E)$ can be calculated using saddle point approximation which gives
\begin{equation}
W(E)=\frac{e^{S}}{(2\pi)^{3/2} {\cal D}^{1/2}}
\label{rhoe}
\end{equation}
where, 
\begin{equation}
S=lnZ+\beta' E - \alpha_p N_p - \alpha_n N_n
\label{s}
\end{equation}
is the entropy with ${\cal Z}$ 
being  the grand canonical partition function in the SPA or SPA+RPA approach. 
The quantities $\alpha_{p,n}$ and $\beta'$
(or $T^{-1}$) are
so chosen that the saddle point conditions
\begin{equation}
E = - \frac{\partial}{\partial \beta'}ln{\cal Z} 
\label{et}
\end{equation}
and
\begin{equation}
N_{p,n}=\frac{\partial}{\partial\alpha_{p,n}}ln{\cal Z}
\end{equation}
are satisfied. The quantity ${\cal D}$ in eq. (7) is a determinant
of $3\times 3$ matrix defined by the elements
\begin{equation}
d_{ij}=\frac{\partial^2}{\partial x_i\partial x_j}
ln {\cal Z}
\end{equation}
where, $x\equiv (\beta',-\alpha_p,-\alpha_n)$ and the second
derivatives are evaluated at saddle points. For the sake of
comparison with the experiment we present below 
the results for $\rho(E)$, instead of $W(E)$, which can be obtained as\cite{Gilbert cjp},
\begin{equation}
\rho(E)=\frac{W(E)}{\sqrt{2\pi\sigma^2}}
\end{equation}
where, $\sigma^2=I^{rig}/\hbar^2$ is the spin cut-off factor with 
 $I^{rig}$ being the rigid - body value of the moment of inertia.

The model space used to perform the numerical calculations is given in 
Table 1. 
For the range of temperature $T\le 2.0$ $MeV$ this basis
space should be adequate. The choice of the values of the spherical single
particle energies is a rather difficult task. Due to the
quantal nature of the nucleus there is no smooth $A$-dependence for
a large range of $A$. From our experience in the
$pf$-shell and rare earth region we have finally chosen these numbers with
the help of Fig. 1 (suitable for $A\approx 100$ ) in Ref. \cite{Skalski npa617}
and Fig. 1 (suitable for $^{108}Sn$) in Ref. \cite{Wads npa559} which
are actually obtained as solutions of Woods-Saxon potential. 
The neutron core with N=40 ensures sufficient number of
active valance neutrons. The value of the core polarization factor,
$f_c$, is taken to be 1.5 . With this the value of the quadrupole
deformation parameter, $\beta$, in the ground state of $^{104}Pd$ comes out
to be about 0.1 which is quite reasonable \cite{dan npa557}. $^{114}Sn$ is spherical in 
the ground state. The pairing correlations are also expected to be
small for these nuclei.
  
For the sake of compactness most of our results and discussions will be
presented for $^{104}Pd$ only. For $^{114}Sn$ results on level densities
will be presented towards the end , before conclusions.
In Fig. 1 we display the SPA as well as SPA$+$RPA
results for the variation of energy as a function of temperature for
$^{104}Pd$ .
To obtain $E(T)$ within the SPA$+$RPA we first of all check the convergence
of the RPA correction factor $\cal {C}_{\scriptscriptstyle {RPA}}$
by choosing various values of $N_m$  in eq. (4). We find that $N_m = 40 $
is sufficient  but we have used $N_m=80$ in the present calculation
so that at very low temperatures it should be sufficiently accurate.
 We see that, as in Refs. \cite{{Lauritzen plb246},{puddu prc47}},
the RPA or quantal fluctuation corrections lower the value of $E(T)$ at
lower temperatures. As temperature increases, quantal fluctuation 
decreases or $\cal {C}_{\scriptscriptstyle {RPA}}\rightarrow 1$ 
yielding the value of $E(T)$ close to the one obtained within
SPA. However, it shows that the RPA corrections are important up to about
$T=2$ MeV.
 We then next show in Fig. 2 the variation of the level
density as a function of $E^*$. The values of $E(0)$ needed
to calculate $E^*$ are obtained in the case of SPA as well as SPA$+$RPA
by extrapolating the corresponding curves of Fig. 1 to $T=0$. These
values come out to be equal to $-310.603$ and $-315.357$ $MeV$ in the 
case of SPA and SPA$+$RPA, respectively. 
 We see that SPA$+$RPA results for the 
level density are in good agreement with the ones extracted recently
\cite{drc prc} through the measurements of proton yields in the reaction
$^{93}Nb(^{12}C, p) ^{104}Pd$. On the other  hand, SPA level densities are 
higher compared to the measured values. For instance, the ratio
$\rho_{\scriptscriptstyle{SPA}}/\rho_{exp}\approx 10^2$ at $E^*=10$ $MeV$ 
which reduces to about 10 at $E^*= 24$ $MeV$. Whereas, 
$\rho_{\scriptscriptstyle{SPA+RPA}}/\rho_{exp}$ varies from 1.0 $-$ 1.2 over
the  entire range of $E^*$ for which experimental data are available.
It should be mentioned that we have not put the error bars on the
experimental data. The experimental data have errors of the order of
20 - 30$\%$ at the lower end and of a few percent at the other end.

Next, we now  discuss about the parameterization of the SPA$+$RPA level
densities in order to extract a value of the level density parameter, $a$.
We have shown in our earlier
work\cite{Agrawal plb339} that SPA level density can be reproduced by Bethe's
formula provided an appropriate value of the parameter '$a$' is used. 
Normally there are two relations which are used to compute the value of $a$,
namely,
\begin{equation}
E^*=a_e T^2
\label{ae}
\end{equation}
and
\begin{equation}
S=2 a_s T
\label{as}
\end{equation}
where, the suffices $e$ and $s$ are used to distinguish the value of $'a'$ 
determined using $E^*$ or  $S$, respectively. Treating $a_e$ and $a_s$
equal, usually a third relation $a=S^2/4E^*$ is also derived. However, in reality
there is no rigorous reason to do so. Usually all the three relations
yield different values of $a$ when computed numerically. Hence, it becomes
 a natural question to ask as to which '$a$' is appropriate for the use
in Bethe's formula for the level densities\cite{Gilbert cjp}:

\begin{equation}
\rho(E)=\frac{\sqrt{\pi}}{12}\frac{e^{2\sqrt{aE^*}}}
{a^{1/4}(E^*)^{5/4}} \frac{1}{\sqrt{2\pi\sigma^2}}.
\label{rb}
\end{equation}

 It is important to realise
from eqs. (7 - 9) that the constant part of the prefactor\cite{Lau prc39}
in eq. (2) which normalizes the 'measure'  does  not 
contribute to  the calculation of $E^*$ but certainly contributes to the
value of $S$. In any mean field approach the effect of this prefactor would
be missing. Now using eqs. (\ref{ae}) and (\ref{as}) we can write 
\begin{equation}
S=2\sqrt{\frac{a_s^2 E^*}{a_e}}=2\sqrt{aE^*}
\label{aes}
\end{equation}
where,
\begin{equation}
a=a_s^2/a_e = S^2/4E^* 
\label{aef}
\end{equation}
We find that the values of $a$ 
required in eq. (\ref{rb}) to get $\rho_{\scriptscriptstyle{SPA+RPA}}$
is quite close to the one given by eq. (\ref{aef}). For instance, at $T = $ 
0.5, 1.0, 1.5 and 2.0 $MeV$ we get $a (a_{fit})$ = 
10.46(11.04), 12.92(13.12), 12.42(12.49), 11.89(11.94), respectively.
In Fig. 3 we have displayed the variation of the inverse level density
parameter ($K=A/a$) as a function of the excitation energy per particle,
$\epsilon = E^* / A$ \cite{Shlomo prc44} which correspond roughly
to $T = $ 0 - 2.0 MeV. The values of $K_e$, $K_s$
and $K_{es}$ are obtained  using $'a'$ calculated from eqs. (\ref{ae}), 
(\ref{as}) and (\ref{aef}), respectively. We see that  in comparison
to $K_e$ and $K_s$ the values of $K_{es}$ are increasing very 
slowly  with $\epsilon$ or temperature for $T \ge 1.0$ $MeV$. Also, we would 
like to point out that the values of
$K_{es}$ are quite close to the experimental values \cite{drc prc} available
for $E^* = 16$ $-$ $22$ $MeV$. Behaviour of $K$ in the very low temperature
region($T \leq 0.5 MeV$)  is reflecting the well known effects of shell
structures.

Finally in Fig. 4 we have displayed
the variation of the level density as a function of $E^*$ for $^{114}Sn$.
The theoretical curves are drawn after dividing the 
actual numbers by a factor 20.827 such that the upper most
point (with minimum uncertainty) of the experimental data
matches with the one calculated in the SPA$+$RPA approach.
The variation is rather very well reproduced by the
solid curve 
keeping in mind the fact that the experimental values have also large
inherent uncertainties \cite{drc prc}. 
The SPA curve is roughly parallel to the SPA$+$RPA one
(the ratio $\rho_{\scriptstyle{SPA}}/\rho_{\scriptstyle{SPA+RPA}} \approx
4.5$ at $E^* = 5.0$ $MeV$ and 2.6 at $E^*=30$ $MeV$)
emplying that here the variation is approximately reproduced.

In conclusion, we have studied the excitation energy dependence of level
densities for $^{104}Pd$ and $^{114}Sn$ including thermal
as well as quantal fluctuations of the nuclear quadrupole shape parameters. 
We find that inclusion of quantal fluctuations is essential to reproduce
the experimental data even upto the excitation energy of about
25 $MeV$ (or $T\approx 1.5$ $MeV$). 
We have also shown that the value of the level density parameter '$a$' to be
used in the Bethe's formula (eq. (\ref{rb})) should be computed from 
eq. (\ref{aef}) only. Also, this value of $a$ can not be used to relate
temperature to get $S$ or $E^*$, particularly in the SPA or
SPA$+$RPA approach. 
Actually the calculation of $a$ from $a_s^2/a_e$ is essentially equivalent to 
fitting the value of the level densities to the Bethe's formula.

\newpage

\begin{table}
\caption { Spherical single-particle energies 
(in units of $\hbar\omega_0$)  with $Z=28$ and $N=40$ 
as a core for proton and neutron, respectively.}
\begin{tabular}{|c|c||c|c|}
\hline
\multicolumn{2}{|c|}{Protons}&
\multicolumn{2}{|c|}{Neutrons}\\
\multicolumn{1}{|c|}{Spherical}&
\multicolumn{1}{|c|}{Energy}&
\multicolumn{1}{|c|}{Spherical}&
\multicolumn{1}{|c|}{Energy}\\
\multicolumn{1}{|c|}{orbits}&
\multicolumn{1}{|c|}{($\hbar\omega_0$)}&
\multicolumn{1}{|c|}{orbits}&
\multicolumn{1}{|c|}{($\hbar\omega_0$)}\\
\hline
$1p_{3/2}$& $-1.376$& $0g_{9/2}$&-0.975\\
$0f_{5/2}$& $-1.374$& $1d_{5/2}$&$-0.484$\\
$1p_{1/2}$&$-1.171 $& $0g_{7/2}$&$-0.30$\\
$0g_{9/2}$& $-0.975$& $2s_{1/2}$&$-0.216$\\
$1d_{5/2}$&$-0.484$& $1d_{3/2}$&$-0.122$\\
$0g_{7/2}$&$-0.30$& $0h_{11/2}$&$-0.122$\\
$2s_{1/2}$&$-0.216$& $0h_{9/2}$&$ 0.358$\\
$1d_{3/2}$&$-0.122$& $1f_{7/2}$&$0.405$\\
$0h_{11/2}$&$-0.122$& $---$& $--$\\
\end{tabular}
\end{table}

\newpage
\begin{figure}
\noindent {\bf Figure Captions}\\
\caption{ Variation of energy as a function of temperature.}

\caption{ Variation of the level density as a function of excitation energy.
Solid circles represent the experimental values taken from Ref. [7].}

\caption {Variation of the inverse level density parameter ($K=A/a$) as a 
function of $\epsilon = E^*/A$. The curves labelled $K_e$, $K_s$ and $K_{es}$ are 
obtained with the values of $'a'$ calculated  using eqs. (13), (14) and 
(17), respectively. }
\caption{Excitation energy dependence of level density for 
$^{114}Sn$. Dashed and Solid curves represent the values of 
SPA and SPA$+$RPA level
density, respectively  reduced by a factor of 20.827 (see the text for details).}
\end{figure}
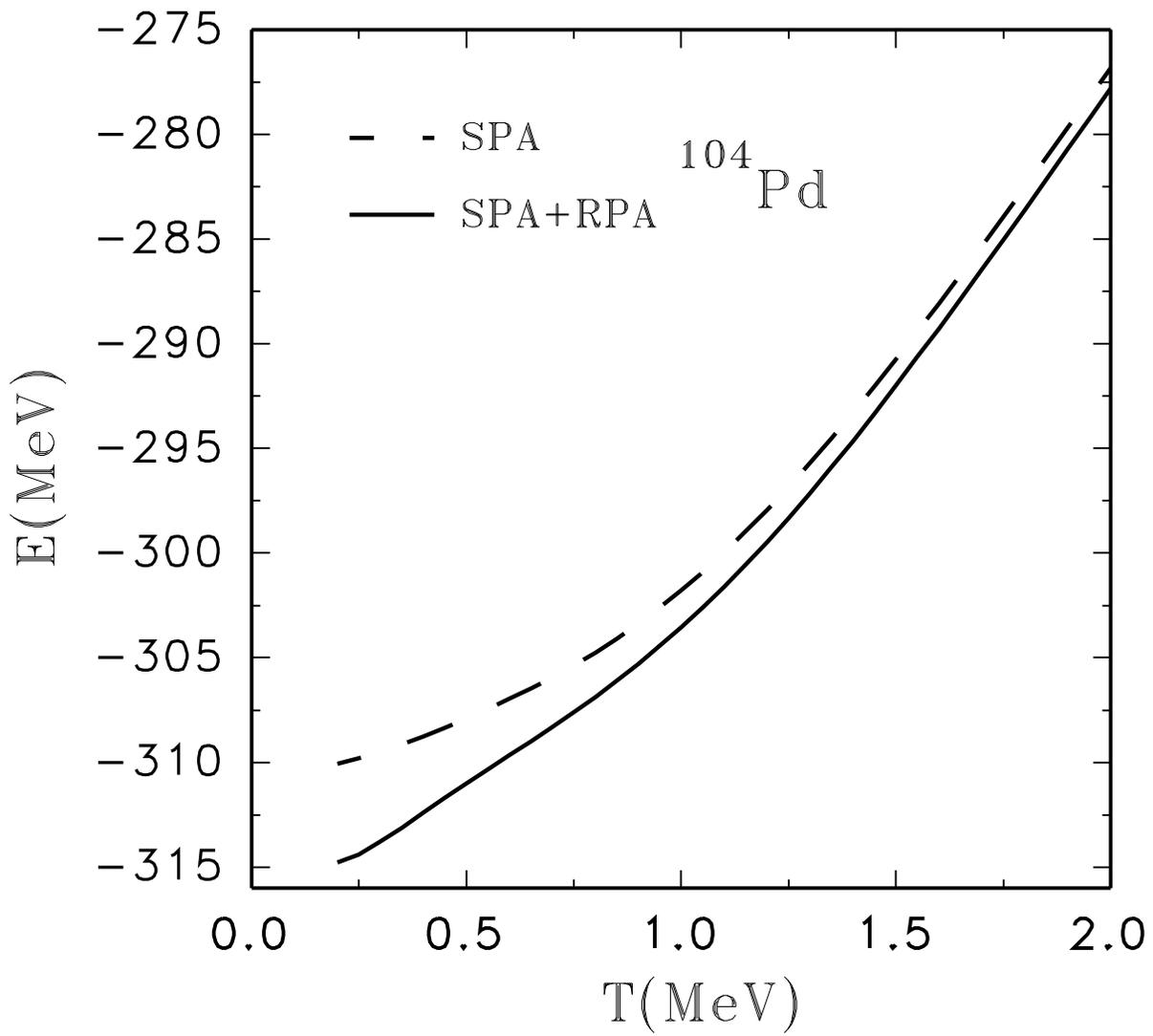
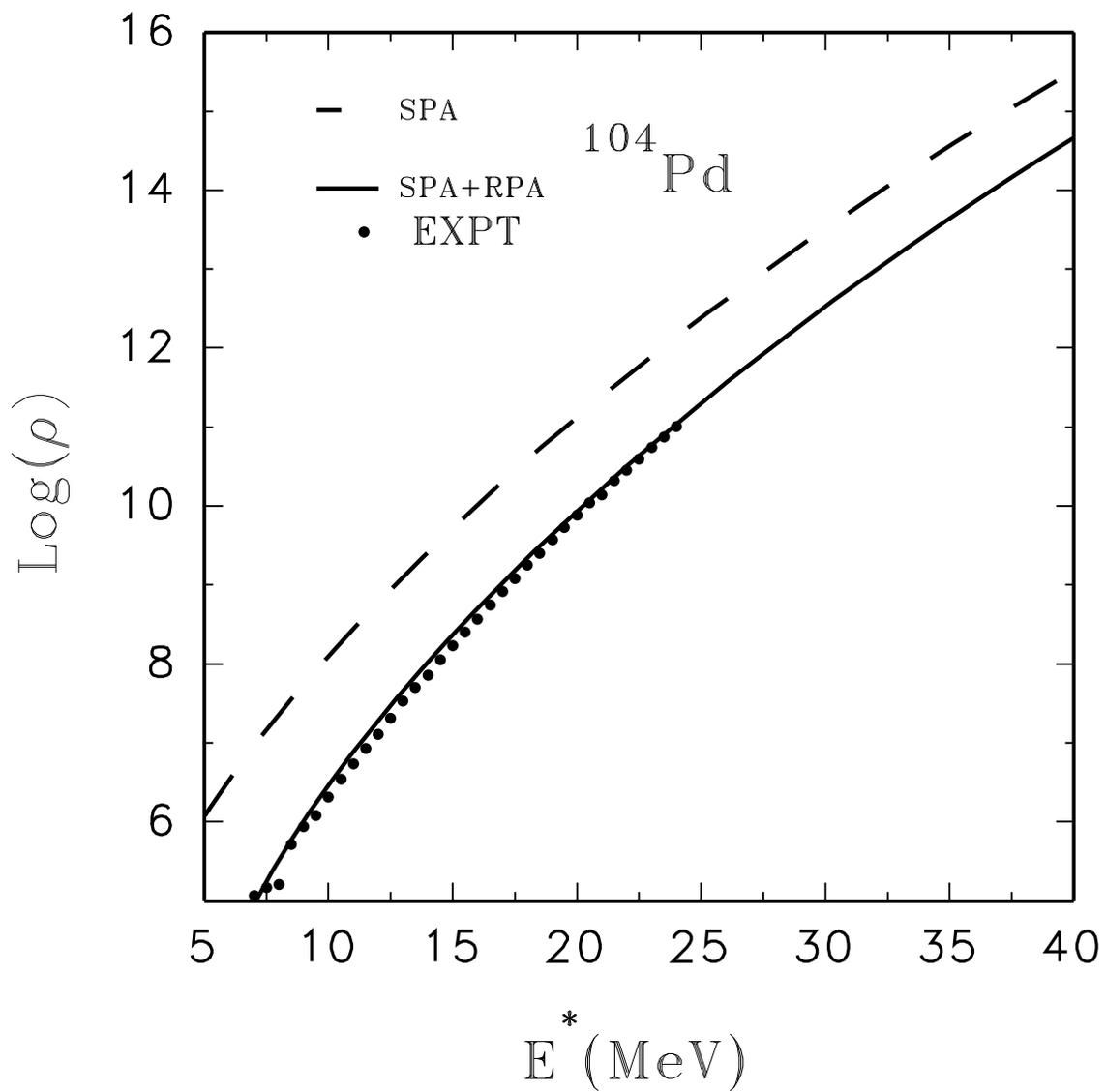
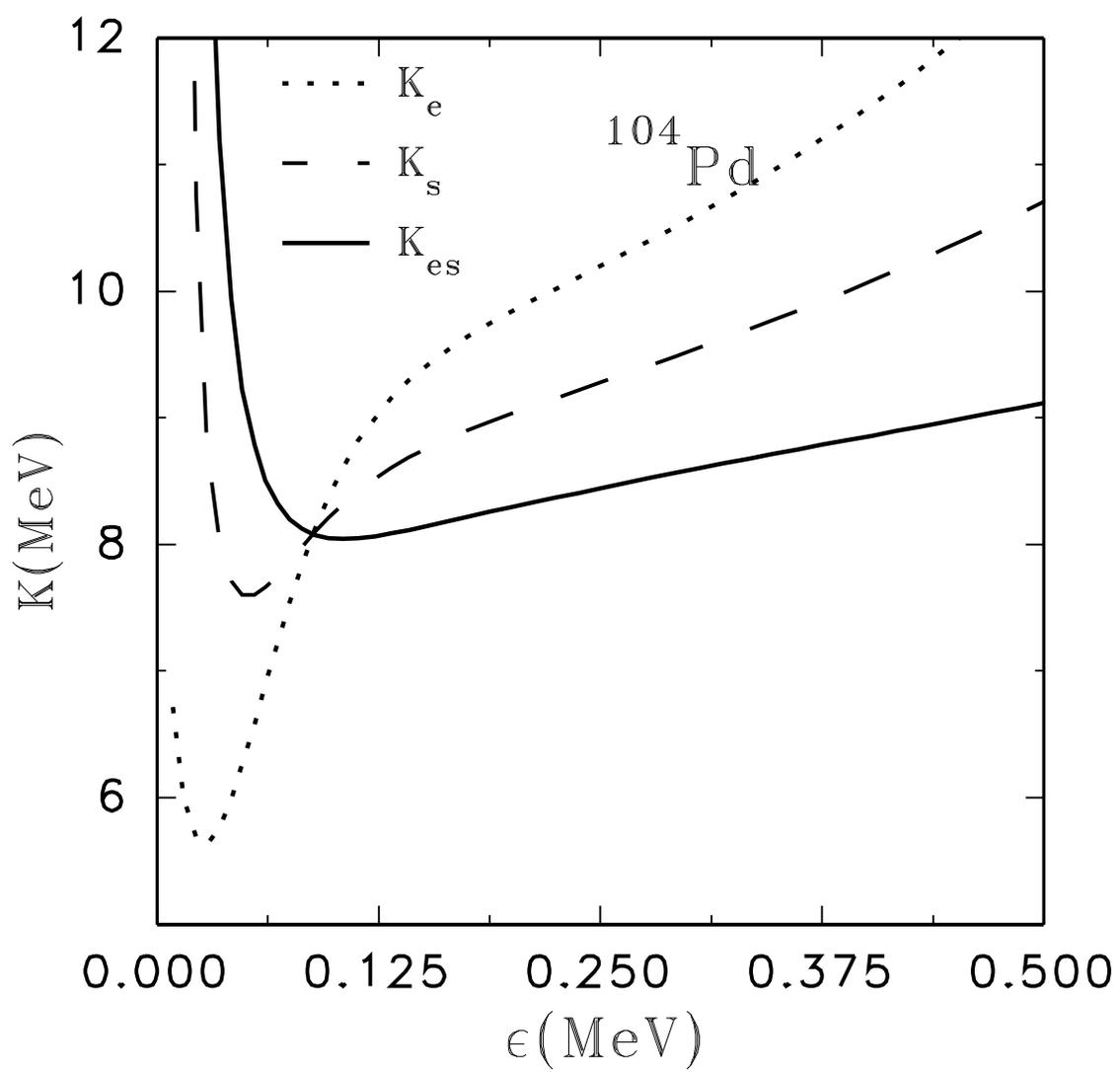
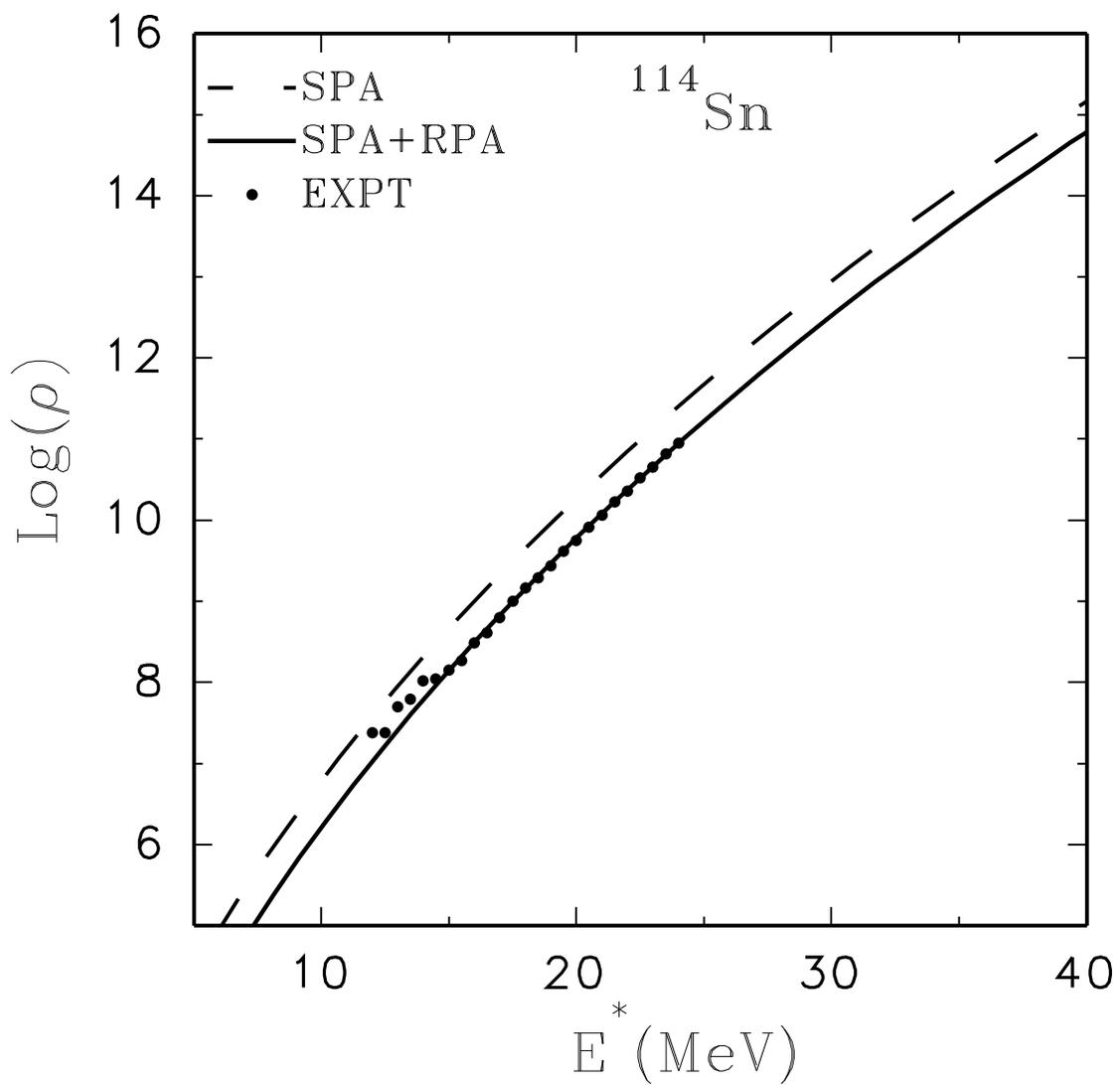
\end{document}